\documentclass[aps, reprint, amsmath, floatfix]{revtex4-1}
\usepackage{amsthm}
\usepackage{mathtools}
\usepackage{xcolor}
\usepackage{graphicx}
\usepackage[left=23mm,right=13mm,top=35mm,columnsep=15pt]{geometry} 
\usepackage{adjustbox}
\usepackage{placeins}
\usepackage[T1]{fontenc}
\usepackage{lipsum}
\usepackage{csquotes}
\usepackage{bm}
\usepackage{comment}
\usepackage[subrefformat=parens]{subcaption}
\begin{document}
\title{A simple theory for training response of deep neural networks}

\author{Kenichi Nakazato}
   \email[Correspondence email address: ] {kenichi_nakazato@sensetime.jp}
    \affiliation{SenseTime Japan Ltd., Minato-ku, Tokyo, Japan}

\date{\today} 

\begin{abstract}
Deep neural networks give us a powerful method to model the training dataset's relationship between input and output. We can regard that as a complex adaptive system consisting of many artificial neurons that work as an adaptive memory as a whole. The network's behavior is training dynamics with a feedback loop from the evaluation of the loss function. We already know the training response can be constant or shows power law-like aging in some ideal situations. However, we still have gaps between those findings and other complex phenomena, like network fragility. To fill the gap, we introduce a very simple network and analyze it. We show the training response consists of some different factors based on training stages, activation functions, or training methods. In addition, we show feature space reduction as an effect of stochastic training dynamics, which can result in network fragility. Finally, we discuss some complex phenomena of deep networks.
\end{abstract}

\keywords{deep neural networks, training response, response decay, response specificity}

\maketitle

\section{Introduction}\label{sec:introduction}
We are in the age of data science, which helps us to predict the future or gather information with much higher quality than in the past. A central issue in the field is statistical inference\cite{statinf1,statinf2,casX}. We want to estimate a value from incomplete information in a statistical way by modeling its statistical features, like statistical physics. Deep neural networks give us a universal model to capture the statistical features of the information, given as a dataset. Neural networks are originally biologically inspired models, which simulate our cognitive process\cite{ANN1,ANN2}. As we learn from experience, the model learns from a dataset. Formally, the model, $f(\bm{x})$, learns the input-output relation in a dataset, $\{(\bm{x}_i,y_i)\}$. By minimizing the error, $|f(\bm{x}_i)-y_i|$,  the network learns the statistical features of the dataset so as to predict correct values.

\textcolor{black}{Neural networks consist of artificial neurons connected with each other and work as an indirect memory, which learns the structural properties in the data space in an adaptive manner as a whole\cite{KBeq}. In other words, it is a typical complex adaptive system, which realizes an adaptive function with the distributed elements\cite{CAS1,CAS2}.} \textcolor{black}{However, we do not fully understand how knowledge is stored, like our own central nervous systems. In fact, we know training results are often not easily reproducible unless we save all of the random factors in the training.}
It is true that we use stochastic optimizing algorithms in large parameter space. That is a source of difficulty in reproducibility but we are still lacking an understanding of training dynamics.

In the studies of such complex systems, we often construct a very simple toy model rather than a real and complicated one. That is because we know a nicely simplified one is enough for understanding the complex ones\cite{CAS1,CAS2}. Actually, stochastic optimization models have been studied in very simplified situations instead of real applications\cite{SOPT}. In such a case, we can derive a clear and general understanding of randomly generated problems and algorithms through analysis of their stochastic nature. An advantage of such an approach is that the results are not dependent on specific problem instances. In other words, we can get to the fundamental nature of the model itself.

Deep neural networks are optimized with a training dataset so as to minimize the target, called a loss function. Usually, that evaluates the error between the pre-defined answer and the output of the network. We can update the network parameters along the gradient of the loss step by step. The training dynamics are generated through the update steps and are often stochastic depending on the update procedures\cite{opts}. Regardless of the stochastic nature, we can derive a deterministic description of the training response at some ensemble levels\cite{TR,ntk1,ntk2}. When we update a network with a data pair of input and output, $(\bm{x}_i,y_i)$, the error is usually reduced on the data point, $\bm{x}_i$, and others, $\bm{x}$, around that. The training response, $\Delta(\bm{x}_i,\bm{x})$, evaluates the reduction of error for any point, $\bm{x}$, by an update step with a specific data point, $\bm{x}_i$\cite{TR}.

\textcolor{black}{If we can assume an infinity limit of the network size, a training response, called the neural tangent kernel or NTK, can be constant in some situations\cite{ntk1,ntk2,ntk3}. Even if we cannot assume the limit, the training response can be described with an almost constant response kernel multiplied by an aging term in an ensemble of dynamics\cite{TR}. It is suggested that the aging follows a power law and the kernel can be varied more or less depending on training stages, however, we do not know the mechanisms behind them. As we know, a power law suggests a universal mechanism, which can be explained with a very simple toy model. We want such a model and understanding of the mechanisms behind the power-law and the kernel shape determination.}

We take an approach to understand the dynamics along the fields of complex systems to study the training dynamics and its mechanisms. Specifically, we construct a very simple network and show that can redisplay the reported phenomena. Then, we analyze the system to elucidate the mechanisms. In the next section, we introduce our model. The model should be simple enough but share a standard architecture in actual applications, like other studies\cite{TR,sgddyn,gandyn}. We thus study a simple and finite-sized network only with a hidden layer. We show that is enough for confirmation of observed phenomena. Then we analyze it with one more simplification.

Finally, we introduce some special cases to show the limit of aging and its result, a fragile network. With our models, we show a few types of aging depending on problem settings and training procedures. At the same time, we elucidate the mechanism behind the variation of the response kernel.
\section{model} \label{sec:model}
As a standard deep neural network, we consider a convolutional neural network\cite{CNN1,CNN2,CNN3,CNN4}. In the network, an input is transformed through multiple convolutions in parallel, and those transformation layers are accumulated into the final output. \textcolor{black}{In each layer, there can be multiple channel inputs, $h_i \ (i=0,\cdots,N_{in})$, and outputs, $h_j \ (j=0,\cdots,N_{out})$,
\begin{eqnarray}
S_j &= \sum_ic_{ij}\circ h_i\\
h_j&=A(S_j+b_j),
\end{eqnarray}
where a nonlinear activation function, $A$, is applied for the transformed input with convolutions, $c_{ij}$, and a bias, $b_j$}\cite{Relu,Elu}.
We usually put one more layer, in which inputs, $\bm{F}$, are transformed into the final output, $y_j$, through a fully-connected linear function,
\begin{equation}
y=A(\bm{a}\cdot\bm{F}+b),
\end{equation}
where the vector, $\bm{F}$, is called a feature vector.
In this way, any input should be encoded into the feature for the last linear evaluation.

As a simplified one, we introduce a network, $f(\bm{x})$, with only one hidden layer.
We assume the input, $\bm{x}$, is a 1-dimensional vector. Even if we have a higher dimensional input, we can transform it into a 1-dimensional one. In a hidden layer, we apply a linear transformation, as a full-size convolution. We get the feature vector through this process,
\begin{equation}
F_i=A(\bm{c}_i\cdot\bm{x}+b_i),\label{eq:Fdyn}
\end{equation}
where we have multiple features through the transformations.
We assume the activation, $A$, is a ramp function, called ReLU.
Finally, the features are transformed into the output,
\begin{equation}
y=\sigma(\sum_i a_iF_i+b).\label{eq:Sdyn}
\end{equation}
As the last activation, we adopt a sigmoid function, $\sigma$.

We have a dataset, ${\cal D}\equiv\{(\bm{x}_i,y_i)\}$, and minimize the error, $L\equiv \sum_i(y_i-f(\bm{x}_i))^2$. As a toy problem, we can consider a random bit encoder or regressor. In the encoder, we transform a 1-dimensional bit string into a binary output. In the regressor, it is transformed into a value, $y\  (0\leq y\leq 1)$. As a symmetric input, we assume each random bit string as a series of +1 and -1.

As a training algorithm, we use an optimizer, known as SGD\cite{opts}. Ideally, this can be regarded as a potential dynamics for each parameter, $w$,
\begin{equation}
\dot{w}\propto-\frac{\partial L}{\partial w}.
\end{equation}
Since we usually have multiple data points in the dataset, corresponding errors, $L_i\equiv L(\bm{x}_i,y_i)$, should be minimized at the same time,
\begin{equation}
\dot{w}\propto-<\frac{\partial L_i}{\partial w}>.
\end{equation}

The training response is defined as.
\begin{eqnarray}
\Delta(\bm{x}_i,\bm{x}) &=& f(\bm{x},\bm{w}+\bm{\delta})-f(\bm{x},\bm{w})\\
&\sim& -\frac{\partial f}{\partial\bm{w}}\cdot\frac{\partial f_i}{\partial \bm{w}}\frac{d L_i}{d f_i}\\
&\equiv&-\Theta(\bm{x}_i,\bm{x})\frac{dL_i}{df_i}.
\end{eqnarray}
The term, $\Theta$, is called a neural tangent kernel, which shows the training effect on the output, $f(\bm{x})$, by the training data point, $(\bm{x}_i,y_i)$\cite{TR,ntk1}.

It is also proposed that the response can be written in an ansatz,
\begin{equation}
\frac{\Delta(\bm{x}_i,\bm{x})}{y_i-y}\propto t^{-\tau}K(\bm{x}_i,\bm{x}),\label{eq:TRansatz}
\end{equation}
where the response kernel, $K$, is almost constant during training and the response decays along a power law with an exponent, $\tau$\cite{TR}. The kernel, $K$, decreases along the distance between the two points, $\bm{x}_i$ and $\bm{x}_i$.
Needless to say, this depiction of training dynamics is based on a linear regime and can be expanded into the case with multiple training inputs easily.

We show our simplified model, (\ref{eq:Fdyn}) and (\ref{eq:Sdyn}), can reproduce the results following the ansatz, (\ref{eq:TRansatz}), and explain the mechanism in the next.

\section{results} \label{sec:results}
\subsection{a simple theory of the training response}
\begin{figure}
\includegraphics[width=1.0\linewidth]{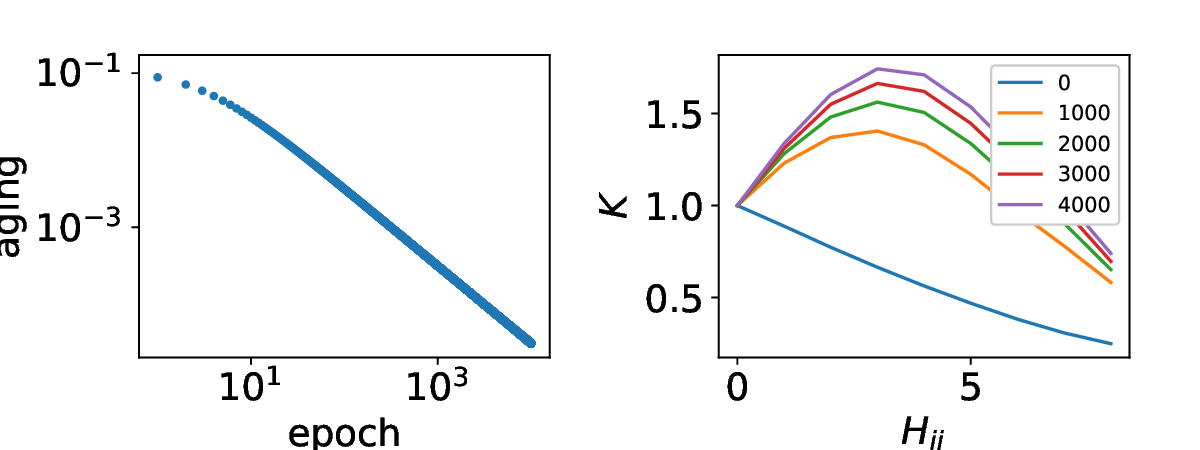}
\caption{Training response with ReLU. The decay of the training response and response kernel are shown on the left and right, respectively. On the left, the training response, $\Delta(\bm{x}_o,\bm{x}_o)$, is plotted against the training epoch. On the right, the response kernel is plotted against the Hamming distance, $|\bm{x}_o-\bm{x}|$. The input size is 8 and the size of the feature is 128.}
\label{fig:reluTR}
\end{figure}

\begin{figure}
\includegraphics[width=1.0\linewidth]{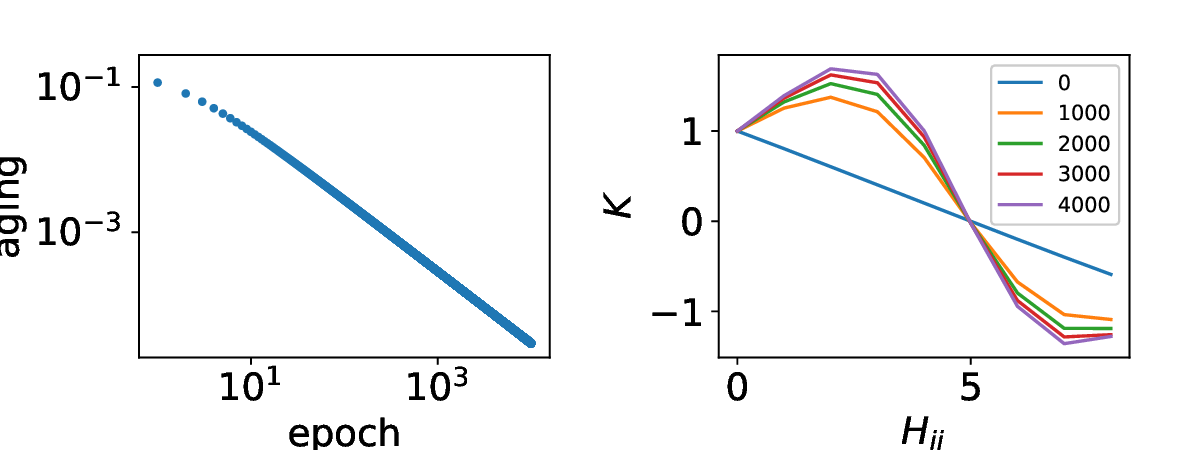}
\caption{Training response with ELU. The settings are the same as FIG. \ref{fig:reluTR}.}
\label{fig:eluTR}
\end{figure}

We show training responses of our model.
In FIG. \ref{fig:reluTR} and \ref{fig:eluTR}, the aging and response kernel are shown on the left and right, respectively. Here, the plots show the results of the training for random bit encoding only with a single training point. We randomly generate a training bit string, $\bm{x}_o$, and its binary encoding, $y_o$, in advance. The same results can be confirmed even if we change the initial condition and the training point. In both of them, we can confirm a power law in the decay of training responses. The response kernels show decreasing tendency against the distance between the training point, $\bm{x}_o$, and the other, $\bm{x}$. However, the tendency changes along the training epochs. The peak gradually shifts away from the point, $\bm{x}_o$.

We can confirm the same power law in both of them, FIG. \ref{fig:reluTR} and \ref{fig:eluTR}. However, the response kernels are somehow different. That of ReLU is positive but that of ELU is not necessarily positive. In addition, S curves emerge in the case of ELU but not in ReLU. We explain the mechanisms behind them later.

\begin{figure}[ht]
  \begin{minipage}[t]{250pt}
\includegraphics[width=1.0\linewidth]{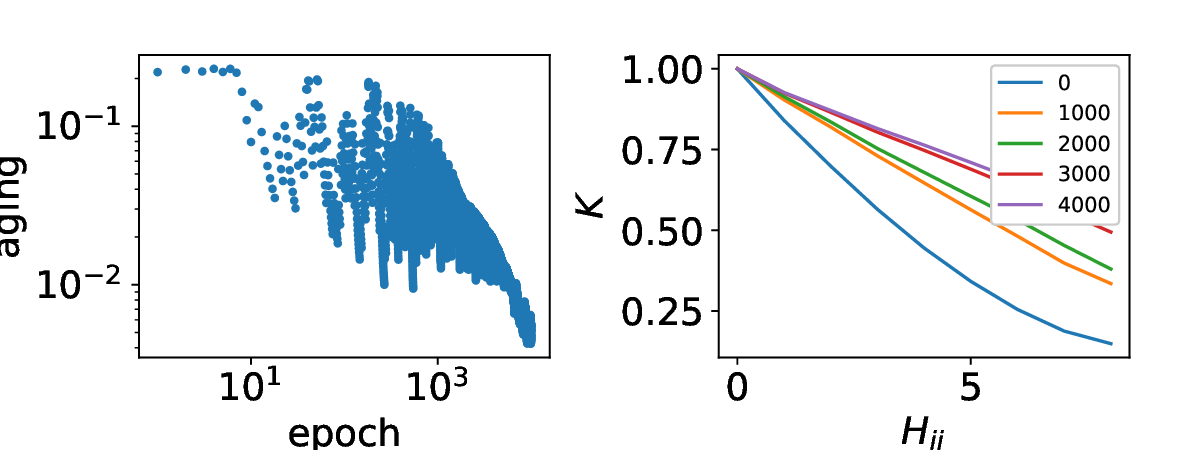}
  \end{minipage}
  \hfill
  \begin{minipage}[t]{250pt}
\includegraphics[width=1.0\linewidth]{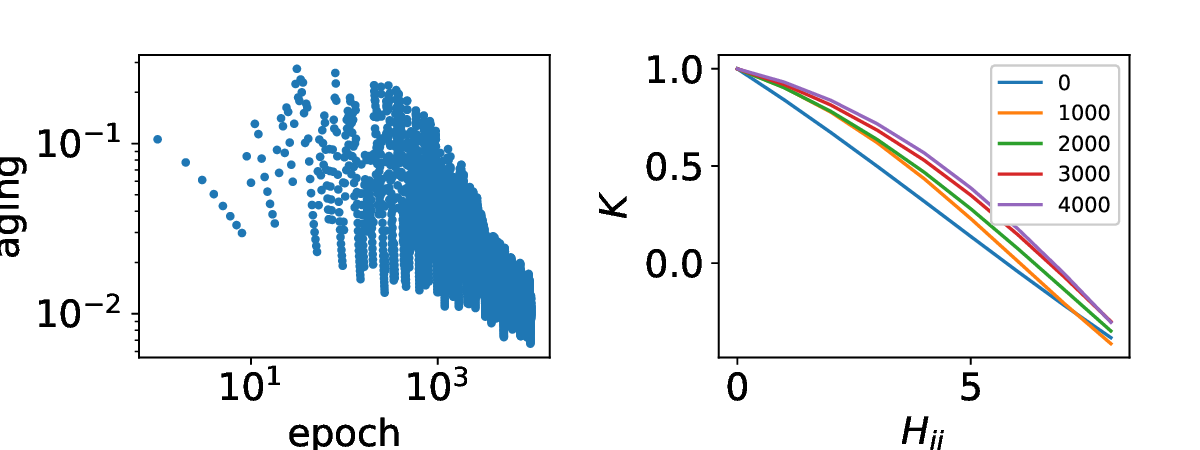} 
\end{minipage}
\caption{The training responses with noise. The training setting is the same one as the FIG. \ref{fig:reluTR}. The upper plots show the results of the training with ReLU. The bottom ones show the results with ELU.}\label{fig:noisedTR}
\end{figure}

We also show the results of training with noise, in FIG. \ref{fig:noisedTR}. The training dataset is a single random bit encoding again but we mix incorrect encoding with a probability, $p=0.1$.
The aging curves are decreasing again, but those do not show a so clear power law, in both cases with ReLU and ELU. In addition the decreasing slopes are different from the former cases, FIG. \ref{fig:reluTR} and \ref{fig:eluTR}. The response kernels are not so variable in comparison to the former ones. We can confirm positive kernels for the case with ReLU but not with ELU, again.

\begin{figure}[ht]
  \begin{minipage}[t]{250pt}
\includegraphics[width=1.0\linewidth]{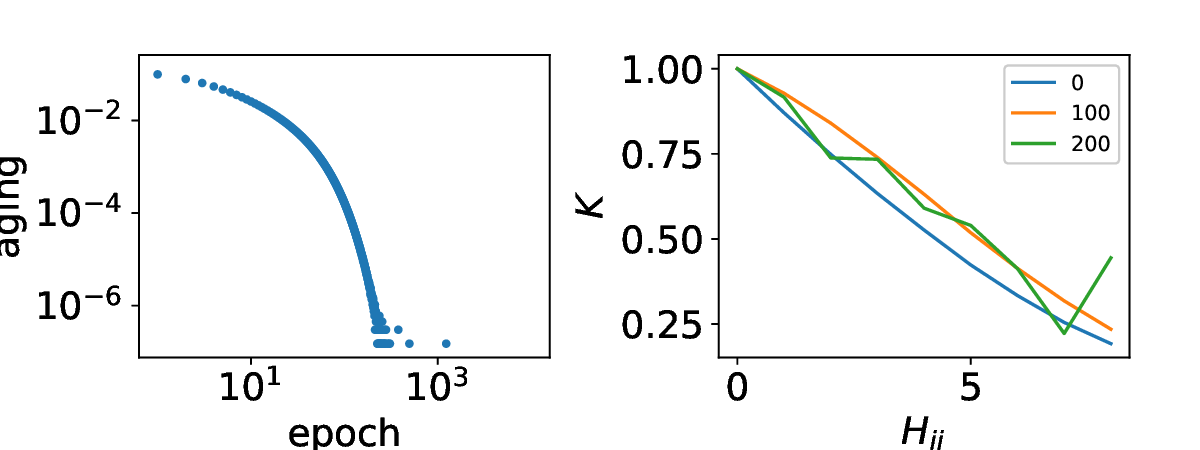}
  \end{minipage}
  \hfill
  \begin{minipage}[t]{250pt}
\includegraphics[width=1.0\linewidth]{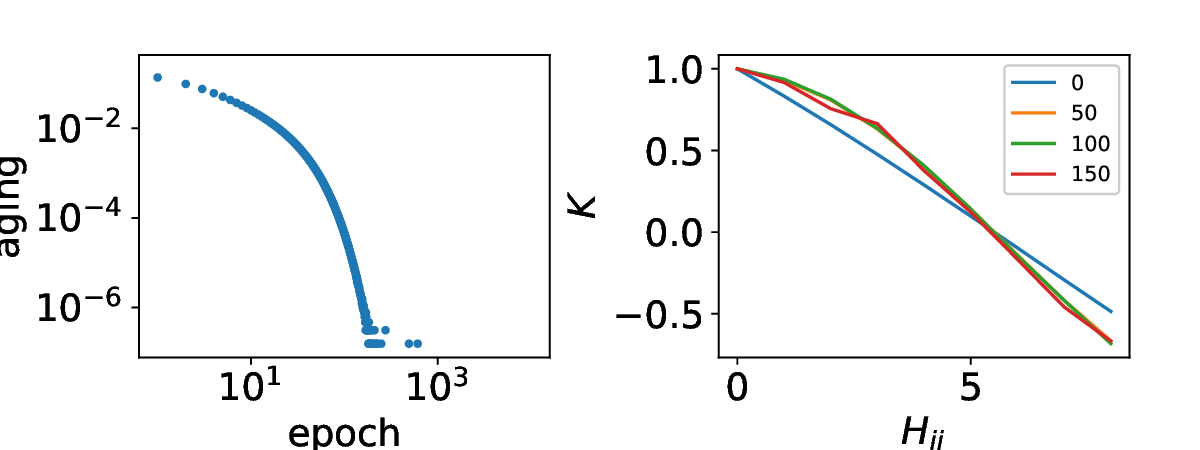} 
\end{minipage}
\caption{The training responses for regression. The setting is the same one as the FIG. \ref{fig:reluTR}, except for the training data point, $(\bm{x}_o,y_o)$, has a different value, $0\leq y_o\leq 1$.}\label{fig:regTR}
\end{figure}

We show one more result, in FIG. \ref{fig:regTR}.  Here, we train the network not for classification but for regression. In other words, the training data point, $(\bm{x}_o,y_o)$, is not binary encoding, $y_o=0\ or\ 1$, but a regression, $0\leq y_o\leq 1$. Different from the others, FIG. \ref{fig:reluTR}, \ref{fig:eluTR} and \ref{fig:noisedTR}, we can observe much faster decay in the training responses and convergence therefore. We only show the response kernels of the earlier stages in the training because of the faster convergence. These results suggest the training response laws are dependent on the problem set.

Next, we derive a training response of our model based on the potential dynamics.
In the model, (\ref{eq:Fdyn}) and (\ref{eq:Sdyn}), we have some parameters to be updated in training, $\bm{c}_i$, $\bm{b_i}$, $a_i$ and $b$. The update of them can be written as follows,
\begin{eqnarray}
\delta \bm{c}_i&\propto&-\frac{\partial L}{\partial \bm{c}_i}\\
&=&(y_o-y)\sigma_o'a_i\bm{x}_o\\
\delta b_i&\propto&(y_o-y)\sigma_o'a_i\\
\delta a_i&\propto&(y_o-y)\sigma_o'F_{io}\\
\delta b&\propto&(y_o-y)\sigma_o'.
\end{eqnarray}
To be noted, we assume the feature, $F_i$, has non-zero gradient in this equations. If the training point, $\bm{x}_o$, is out of sensitive region, $\bm{c}_i\cdot\bm{x}_o+b_i<0$, those terms can be ignored.
On the other hand, the update of the output, $y$, can be written,
\begin{eqnarray}
\delta y&=&\sigma'(\sum\delta a_iF_i +\sum a_i\delta F_i +\delta b)\\
&=&\sigma'(\sum \delta a_iF_i + \sum a_i(\delta \bm{c}_i\cdot\bm{x}+\delta b_i) +\delta b).
\end{eqnarray}
Therefore, we can write the training response,
\begin{equation}
\frac{\delta y}{y_o-y}\propto\sigma'\sigma_o'(\sum(F_{io}F_i+a_i^2(\bm{x}\cdot\bm{x}_o+1))+1).
\end{equation}

If we can assume the parameters are almost constant anymore in training, the equation can be simplified more,
\begin{equation}
\frac{\delta y}{y_o-y}=\sigma'\sigma_o'(\sum C_0+C_1\bm{x}\cdot\bm{x}_o+C_2),\label{eq:simpleTR}
\end{equation}
where we rewrite it with the constants, $C_0$, $C_1$ and $C_2$.
After enough training, the gradient of the activation, $\sigma_o'$, is very small but the other is still larger than that, in other words, $\sigma_o'\sigma_o'<\sigma'\sigma_o'$. That is why we observe the shift of peaks in the response kernels, in FIG. \ref{fig:reluTR} and \ref{fig:eluTR}. Furthermore, the term, $\bm{x}\cdot\bm{x}_o$, decreases along the increase of the distance between them. The dot product can be negative, specifically when $\bm{x}\sim-\bm{x}_o$. However, such a negative response can be cancelled when we use the positive activation, ReLU.

\begin{figure}[ht]
  \begin{minipage}[b]{0.45\linewidth}
\includegraphics[width=1\linewidth]{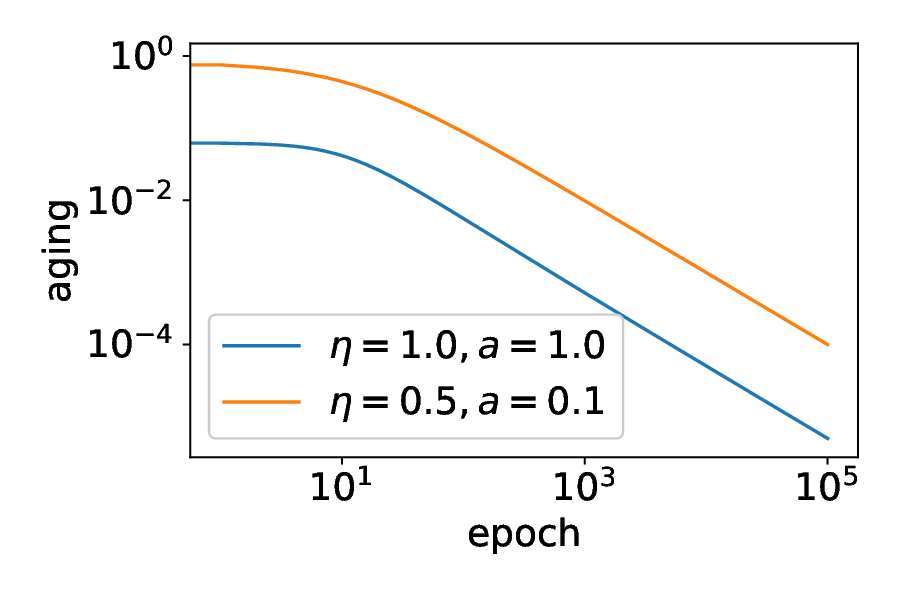}
  \end{minipage}
  \begin{minipage}[b]{0.45\linewidth}
\includegraphics[width=1\linewidth]{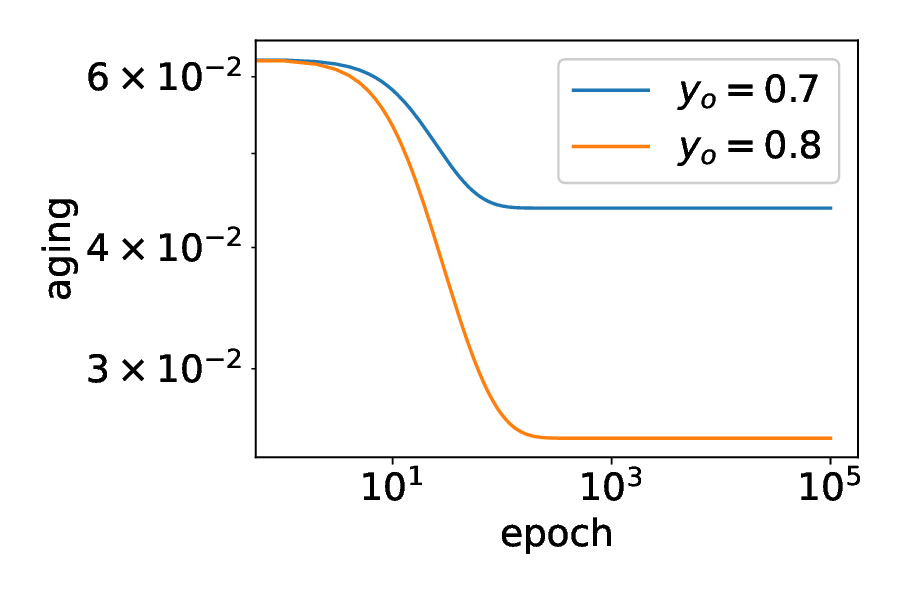} 
\end{minipage}
\caption{The training responses with a simplified system. On the left, we show results with a target value, $y_o=1$, and different training rates, $\eta$, and slope parameters, $a$. On the right, we show results with two target values, $0<y_o<1.0$. Training rate and slope parameters are the same one, $\eta=1.0$ and $a=1.0$.}\label{fig:simpleA}
\end{figure}

Since the aging is the training response at the point, $\bm{x}_o$, we can write down it in a simpler form,
\begin{eqnarray}
\frac{\delta y}{y_o-y}&=&C\sigma'(x)^2\label{eq:TRform1}\\
\delta x&=&\eta(y_o-y)\sigma'(x)\label{eq:TRform2},
\end{eqnarray}
where the term, $C$, is a constant. We can show this system generates a power law-like behavior in a wide condition on the parameters and the activation function. However, the aging shows much faster decay to the constant response when the target value, $y_o$, is not a binary one. This means the aging curves are dependent on a relation between the activation function and the target value, in other words, problem type, shown in FIG.\ref{fig:simpleA}.

\subsection{feature space reduction}
\begin{figure}[ht]
  \begin{minipage}[b]{0.45\linewidth}
\includegraphics[width=1\linewidth]{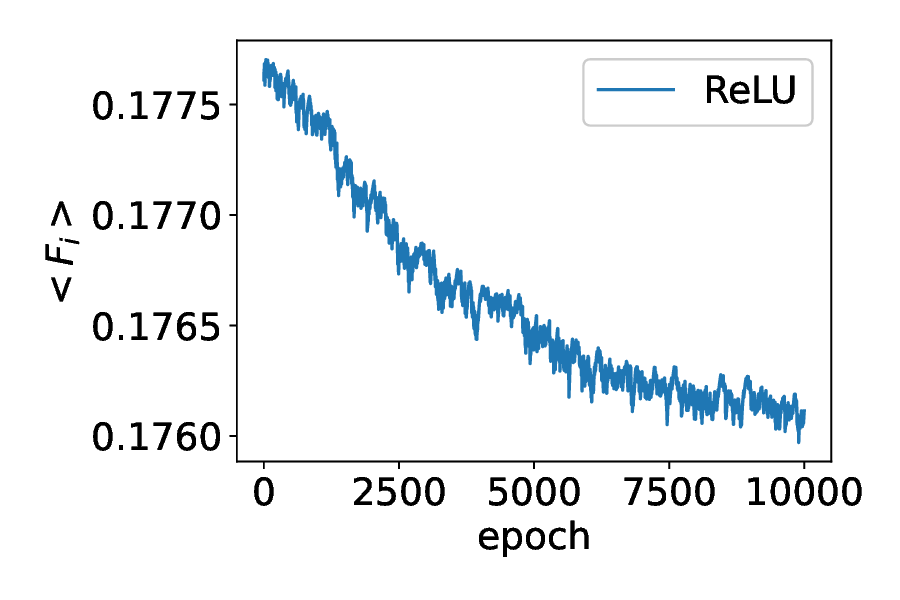}
  \end{minipage}
  \begin{minipage}[b]{0.45\linewidth}
\includegraphics[width=1\linewidth]{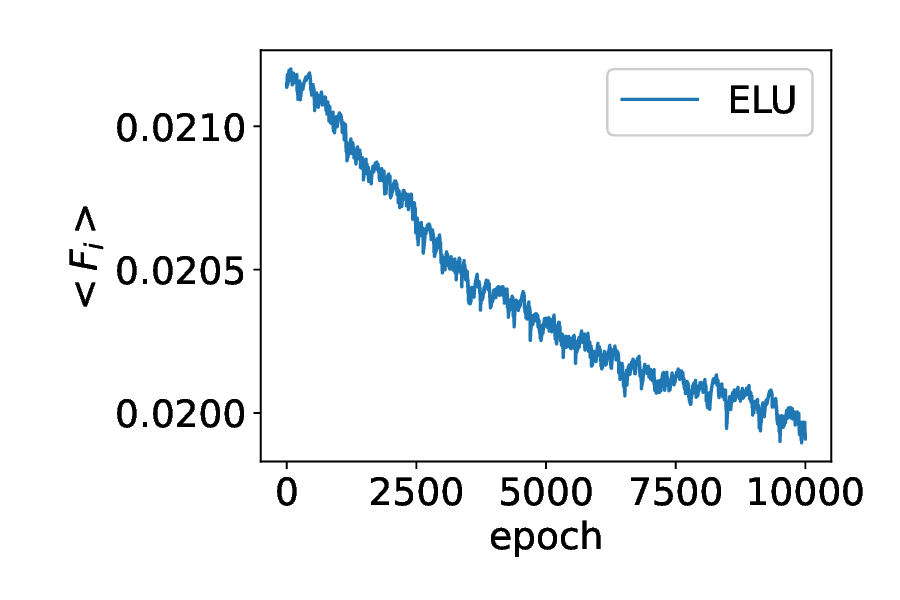} 
\end{minipage}
\caption{Feature space reduction. The averaged value of feature vectors, $<F_i>$, are plotted with different activation functions. The parameters are the same with the ones in FIG. \ref{fig:reluTR}.}\label{fig:features}
\end{figure}

In the case of random training, in FIG. \ref{fig:noisedTR}, the prediction, $f(\bm{x}_o)$, can not converge into a binary value but we have an expected value, $0<E(y_i)<1$. Since the activation function, $\sigma$, is not flat in such a region, we cannot expect a power law decay with the described mechanism. However, we can confirm continuous aging, in FIG. \ref{fig:noisedTR}, different from the regression patterns, in FIG. \ref{fig:regTR}. For the explanation of the other types of aging, we show one more results, in FIG. \ref{fig:features}. We can confirm the reduction of feature space. On the contrary, we can show the increasing curves in training without noise.
Since the activation functions, ReLU and ELU, have an almost gradientless region, features can be irreversibly lost in the course of training. This results in the reduction of feature space and aging, therefore.

\begin{figure}[ht]
  \begin{minipage}[b]{0.45\linewidth}
\includegraphics[width=1.0\linewidth]{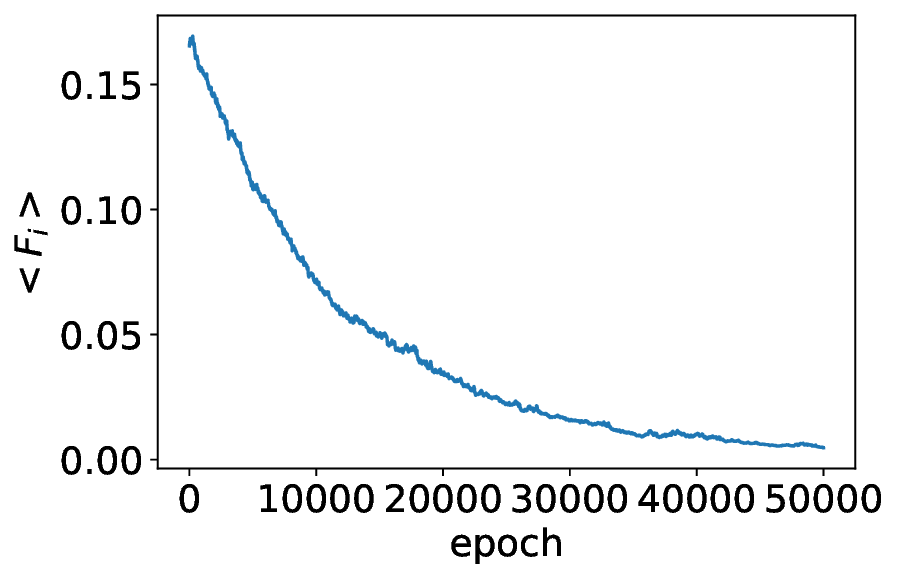}
  \end{minipage}
  \begin{minipage}[b]{0.45\linewidth}
\includegraphics[width=1.0\linewidth]{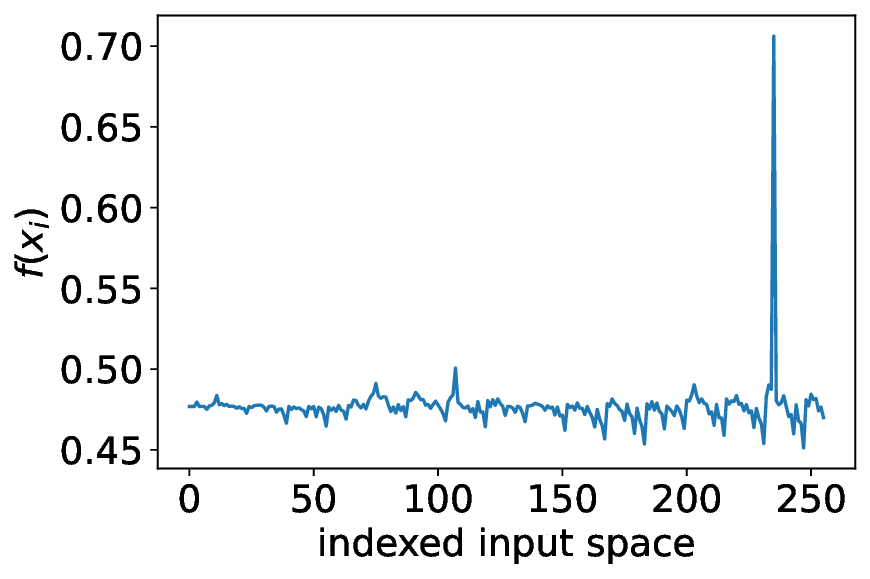} 
\end{minipage}
\caption{The results of random training. On the left, the averaged feature is plotted. As the training dataset, a randomly generated bit string, $\bm{x}_i$, and a binary value, $y_i$, are given at every training epoch. On the right, the prediction, $f(\bm{x}_i)$, is plotted. We prepared a fixed pair, $(\bm{x}_o,y_o)$, in advance. In this case, we used a value, $y_o=1$. In addition, a randomly generated pair is replaced with that with a probability, $p=0.2$, in each epoch. The activation function is ReLU in both cases. The other setting is the same as FIG. \ref{fig:reluTR}.}\label{fig:randomT}
\end{figure}

We can show the feature space reduction more clearly with random training. In the case, FIG. \ref{fig:features}, the model is trained with a mixed dataset, $(\bm{x}_i,y_i)$ and $(\bm{x}_i,-y_i)$. Here we show two more results of such random training. We can use pairs of randomly generated bit string, $\bm{x}_r$, and a binary, $y_r$, at each training epoch. We call it as random training, here. We can show the feature space is lost by such random training, on the left of FIG. \ref{fig:randomT}. In addition, we add a fixed pair, $(\bm{x}_o,y_o)$, to random training, shown in the right of FIG. \ref{fig:randomT}. We show the prediction of the trained model after 50000 epochs. As we can confirm, the model learns a delta function, $\delta(\bm{x}-\bm{x}_o)$, finally. Almost all features are lost, except for ones which have a specific sensitivity against the input, $\bm{x}_o$.

\section{discussion}\label{sec:discussion}

In a few decades, statistical inference has attracted much interest in the field of not only computer science but also statistical physics. Actually, many studies have been reported on the relation between statistical inference and the Ising model\cite{statinf2}. Among them, an unsupervised neural network, the Boltzmann machine, is a stochastic spin-glass\cite{BM1,BM2}. In other words, neural networks can have complex behavior in some situations. On the contrary, deep neural networks are usually defined as supervised training literature and are not so similar to the Ising model. However, its complex behavior has been suggested\cite{adversarial1,adversarial2,TR,ntk3}. Deep networks can memorize complicated relations in the dataset, ${\cal D}\equiv\{(\bm{x}_i,\bm{y}_i)\}$, and predict for any input, $\bm{y}=f(\bm{x})$, in a high precision. This feature is known as a generalization but it is often fragile, $f(\bm{x})\neq f(\bm{x}+\bm{\delta})$\cite{adversarial1,adversarial2}. As we can see in the training ansatz, (\ref{eq:TRansatz}), such a discontinuity in the output is not so natural. However, deep networks often show such fragility against input perturbation. \textcolor{black}{In addition, training dynamics is not so straightforward and often not easily reproducible, as we know.} Deep networks are trained for minimizing the defined loss function through potential dynamics. This means its convergence depends on a learning rate and landscape complexity\cite{convergence_issue1,convergence_issue2}. On the other hand, it is known that training dynamics can be described with the training response in a simple form in some cases\cite{ntk1,ntk2,TR}. Here, we study a very simple network for understanding the mechanisms behind the simple form and such complex behaviors.

\textcolor{black}{The training response can be written in a simple ansatz, (\ref{eq:TRansatz}), consisting of aging term and the response kernel\cite{TR}. Our model, (\ref{eq:Fdyn}) and (\ref{eq:Sdyn}), is enough to reproduce the ansatz, shown in FIG. \ref{fig:reluTR} and \ref{fig:eluTR}. However, the response kernel gradually varies its shape along training epochs. As we have shown, the training response of our model can be simply expressed in an equation, (\ref{eq:simpleTR}),  which explains the variation of the response kernel, as we wrote in the previous section. The evolution of aging can be written in a system, (\ref{eq:TRform1}) and (\ref{eq:TRform2}), in the same way. As we have shown, it reproduces the behaviors of aging, in FIG. \ref{fig:simpleA}. The training response is almost decreasing by the term, $\bm{x}\cdot\bm{x}_o$, in our model but it can change its form depending on the activation function. In addition, the feature of aging also depends on the shape of the activation function. Specifically, if the target value is not in the flat region in the function, we cannot observe power law-like behavior. In other words, the power law does not require an infinity limit in its size. Instead, a very simple network is enough for the understanding.
Our findings suggest more complex neural networks can be reduced into our simple model, if our interest is the training response.}

Furthermore, we found one more mechanism behind aging. The features can be lost by random training, in FIG. \ref{fig:features} and \ref{fig:randomT}. As we commented, the training response can be reduced by such a loss of features. As specific cases, we can show a response-less network and delta function network by random training, in FIG. \ref{fig:randomT}. As we notice, the delta function means not generalized but fragile prediction, which can be changed in a discontinuous manner for the perturbed inputs. In a more complicated network, we can have response-less neurons here and there for some reasons. The sensitivity of the neurons can be lost by accident, in other words, the stochastic nature of training or initial parameters. Even if we cannot observe such a discontinuous prediction within the training dataset, irregular responses of neurons can be activated for unknown inputs. In fact, it is consistent with the report, which says smoothed activation functions can be effective to enhance adversarial robustness\cite{smoothed_adv}. Our theory gives an explanation for such an improvement.

Our theory explains how the training response changes along the situations. It is reported that NTK can be constant in a limit\cite{ntk1,ntk2}. There, NTK can be derived with many hidden units and linearization. Needless to say, our derivation is consistent with the findings. Our expression fills the gap between the constant NTK and the ansatz for the training response in a non-linear regime\cite{TR}. The response kernel shows the interpretation, $K(\bm{x}_o,\bm{x})$, of the training data, $(\bm{x}_o,y_o)$, in the training effect. As the kernel shows, the network output is optimized not only for the point, $\bm{x}_o$, but also for the other, $\bm{x}$, along with the weight, $K(\bm{x}_o,\bm{x})$. This is the mechanism behind generalization. However, as our theory shows, features or neurons are divided into two groups, active or not to the point, $\bm{x}_o$. This feature of nonlinear function can enhance the flexibility of the network for better or worse. We found the activity can be varied as a mechanism of aging. This suggests the model complexity is reduced through aging. In other words, over-fitting may be reduced by the inactivation of neurons. However, it is not easy to control the inactivation because that is effectively irreversible and results in nonequilibrium dynamics. In the real world, the dataset is not necessarily fixed but somehow open. The dataset is revisioned, updated, or replaced with new ones under the data stream, like our experience. Since the neural networks can be irreversibly damaged by random training, the consistency of the dataset is necessary for maintenance.

The equation, (\ref{eq:TRansatz}), shows the training response for training with one data point, $(\bm{x}_i,y_i)$. Our results also concerns about the one point training. If we have multiple data point in training, the response consists of the sum of them. Given the network can be reduced into our simple one, (\ref{eq:Fdyn}) and (\ref{eq:Sdyn}), the total response can be understood as the weighted average with the weight, $y_i-y$. If the weights are significantly different, the aging would not be so straightforward. However, the short-term training dynamics in earlier stage can be understood enough with the description of training response and our theory. Our toy model approach assures the reproduction of the macroscopic law of training response. This suggests the toy model is enough for understanding even more complicated networks, as we mentioned.

In general, deep neural networks are useful for cases with large datasets. As our model shows in the response kernel, (\ref{eq:simpleTR}), the networks tend to output similar predictions for any inputs, $\bm{x}\sim\bm{x}_i$, along the kernel. Needless to say, we need a more dense dataset if we assume a complicated structure in the data space. In addition, we need much more training epochs for higher accuracy in such a case, as shown in reports\cite{TR,scalingLLM}.

As further works, we note some points. In the field of statistical physics, we usually rely on Monte Carlo algorithms for optimization, like simulated annealing\cite{SA}. Some studies have been reported on the line\cite{SAdnn,aMC}. However, it is not so easy and requires some modifications\cite{aMC}. Specifically, it fails when the network is highly heterogeneous. The training response, (\ref{eq:simpleTR}), seemingly does not have the source of heterogeneity but we can have lost features by the nonlinearity of the activation function. In fact, the response can be heterogenous, because of the nonlinearity, when we have multiple training inputs. We need more understanding on the heterogeneity of the landscape and its relation to the network structure.

We know the network skeleton is often effective to understand the complex networks\cite{skeleton}.
In fact, the existence of the network skeleton has been discussed and the sub-network may be easy to be optimized\cite{lottery1,lottery2}. In applications, we often apply pruning for the network so as to minimize the network. However, the precision can be kept in some ways\cite{pruning}. In our model, some of the features can be removed without loss of precision. This suggests we have good initial conditions suited for the reduced model, known as the lottery ticket hypothesis.

As one more issue on the landscape structure, we do not know much about the capacity of the deep networks yet. We can derive the critical capacity in a simple perceptron\cite{statmechML1,statmechML2,annrevML}. However, we still need more practical understanding between the size of dataset, the model complexity, training epochs, and the error. We know training dynamics based on the ansatz, (\ref{eq:TRansatz}), is not straightforward depending on the complexity of the data structure\cite{TR}. On the other hand, as an empirical law, we know simple scaling rules on those values can be applied for very large networks\cite{scalingLLM}. 

\textcolor{black}{As we know, a power law suggests a universal underlining mechanism, which can be explained in a very simple model. In the field of network science, there have been many reports on complex networks\cite{complex-net}. A famous power-law mechanism is preferential attachment, which realizes the power law in its degree distribution\cite{BA}. Actually, such a power law distribution can be observed in artificial neural networks\cite{PA-NN}. Such a structural aspect of the networks is out of our focus, but the relation with the training dynamics is an interesting future topic.
}

\textcolor{black}{We often compare machine learning algorithms to show the superiority of the new one in proposals. On the other hand, in this paper, we compare the experimental results with a new theory as an explanation or understanding. In this meaning, our contribution to the field is not the new algorithm but the theoretical explanation of the training dynamics. We hope that our theory leads us to new technologies and high-performance algorithms as future contributions. In other words, we should fill the gap between those complex phenomena and our understanding. Needless to say, we believe statistical and dynamical analysis for very simple toy models must be effective, like other fields of complex systems.}

\section*{Acknowledgements} \label{sec:acknowledgements}
This paper is motivated through works with SenseTime Japan, BCAI and other former colleagues, MT and HK.


\end{document}